\documentclass[
    final            % use final for the camera ready runs
  ]
  {aipproc}

\layoutstyle{6x9}

\begin{document}

\title{Strangeness contribution to the vector and axial form
factors of the nucleon}



\author{Stephen Pate}{
  address={Physics Department, New Mexico State University, Las Cruces NM 88003, USA}
}

\author{Glen MacLachlan}{
  address={Physics Department, New Mexico State University, Las Cruces NM 88003, USA}
}

\author{David McKee}{
  address={Physics Department, New Mexico State University, Las Cruces NM 88003, USA}
}

\author{Vassili Papavassiliou}{
  address={Physics Department, New Mexico State University, Las Cruces NM 88003, USA}
}

\begin{abstract}
The strangeness contribution to the vector and axial form
factors of the nucleon is presented for momentum transfers in the
range $0.45<Q^2<1.0$ GeV$^2$.  The results are obtained via a combined
analysis of forward-scattering parity-violating elastic $\vec{e}p$
asymmetry data from the $G^0$ and HAPPEx experiments at Jefferson Lab, and elastic
$\nu p$ and $\bar{\nu} p$ scattering data from Experiment 734 at
Brookhaven National Laboratory.  The parity-violating asymmetries
measured in elastic $\vec{e}p$ scattering at forward angles establish
a relationship between the strange vector form factors
$G_E^s$ and $G_M^s$, with little sensitivity to the strange axial form
factor $G_A^s$.  On the other hand, elastic neutrino scattering at low
$Q^2$ is dominated by the axial form factor, with still some
significant sensitivity to the vector form factors as well.  The
combination of the two data sets allows the simultaneous extraction of
$G_E^s$, $G_M^s$, and $G_A^s$ over a significant range of $Q^2$ for
the very first time.
\end{abstract}

\maketitle

%%%%%%%%%%%%%%%%%%%%%%%%%%%%%%%%%%%%%%%%%%%%
%% MAINMATTER
%%%%%%%%%%%%%%%%%%%%%%%%%%%%%%%%%%%%%%%%%%%%

{\bf Electroweak Currents and the Elastic Form Factors of the Nucleon:}
The static properties of the nucleon are described by elastic form factors
defined in terms of matrix elements of current operators.
The electromagnetic (one $\gamma$ exchange) interaction involves two
vector operators and thus two vector form factors,
$G_E^{\gamma,N}(Q^2)$ and $G_M^{\gamma,N}(Q^2)$,
where $Q^2=-(p-p')^2$ is the momentum transfer between two nucleon states,
and $N$ is for proton ($p$) or neutron ($n$).
Similarly, the neutral weak current (one $Z$ exchange) involves two analogous
vector form factors $G_E^{Z,N}(Q^2)$ and $G_M^{Z,N}(Q^2)$ and also an
axial form factor $G_A^{Z,N}(Q^2)$.
Due to the point-like interaction between the gauge bosons ($\gamma$ or $Z$)
and the quarks internal to the nucleon, these form factors can be expressed
as separated contributions from each quark flavor; for example, the electromagnetic
and neutral weak electric form factors of the proton 
can be expressed in terms of up, down, and strange quark form factors:
\begin{eqnarray*}
G_E^{\gamma,p} &=& \frac{2}{3}G_E^u - \frac{1}{3}G_E^d - \frac{1}{3}G_E^s \\
G_E^{Z,p} &=& \left(1-\frac{8}{3}\sin^2\theta_W\right)G_E^u
+\left(-1+\frac{4}{3}\sin^2\theta_W\right)G_E^d
+\left(-1+\frac{4}{3}\sin^2\theta_W\right)G_E^s.
\end{eqnarray*}
The same quark form factors are involved in both expressions; the 
coupling constants that multiply them (electric or weak charges) correspond
to the interaction involved (electromagmetic or weak neutral current).
These measurements are most interesting for low momentum 
transfers, $Q^2< 1.0$~GeV$^2$, as the $Q^2=0$ values of these form factors
represent static integral properties of the nucleon.  Of most significance here
is the fact that the $Q^2=0$ value of $G_A^s$ is the strange quark contribution
to the nucleon spin, $\Delta s$, which is also the first moment of the 
polarized strange quark momentum distribution $\Delta s(x)$ measured in deep-inelastic
scattering:  $\Delta s = G_A^s(Q^2=0) = \int_0^1 \Delta s(x) dx$.

{\bf Parity-violating forward scattering {\em ep} data:}
Several experiments\footnote{See talks by Doug Beck and Jianglai Liu.} 
have now produced data on forward PV $\vec{e}p$ elastic
scattering~\cite{Aniol:2004hp,Maas:2004dh,HAPPEx_1H_010,Armstrong:2005hs,
Maas:2003xp}.  Of most interest here are measurements that lie in the same $Q^2$ range
as the BNL E734 experiment, which are the original HAPPEx measurement~\cite{Aniol:2004hp}
at $Q^2=0.477$~GeV$^2$ and four points in the recent $G^0$ 
data~\cite{Armstrong:2005hs}.  These forward scattering data are most sensitive to $G_E^s$,
somewhat less sensitive to $G_M^s$, and almost completely insensitive to the axial form 
factors due to supression by both the weak vector electron charge $(1-4\sin^2\theta_W)$
and by a kinematic factor that approaches 0 at forward angles.  Thus, the forward
PV $ep$ data
do not provide information directly about the strange axial form factor, but
provide an important contraint on the vector form factors which is needed to make
the neutrino data useful.

{\bf Elastic neutrino-proton data:}
The world's only data on elastic $\nu p$ and $\bar{\nu}p$ scattering comes
from the BNL E734 experiment~\cite{Ahrens:1987xe}, and cover the range
$0.45<Q^2<1.05$~GeV$^2$.  Due to a variety of experimental and analytical
difficulties, these data have large total uncertainties, typically 20-25\%.  
These data are primarily
sensitive to the axial form factor of the proton --- the axial contribution dominates
at low $Q^2$.  However, knowledge of the strange vector form factors is still necessary
for a clean extraction of the strange axial form factor from these data, as was demonstrated
in Ref.~\cite{Pate:2003rk}.  Previous analyses of these 
data~\cite{Ahrens:1987xe,Garvey:1993cg,Alberico:1998qw} 
were hampered by a lack of knowledge of the strange vector form factors and so had to assume
a form for the $Q^2$-dependence of $G_A^s$ --- no such assumption is necessary any longer, now
that the PV $ep$ data are available to constrain the strange vector form factors.

\begin{figure}[t]
  \includegraphics[height=.48\textheight, bb=150 195 495 570]{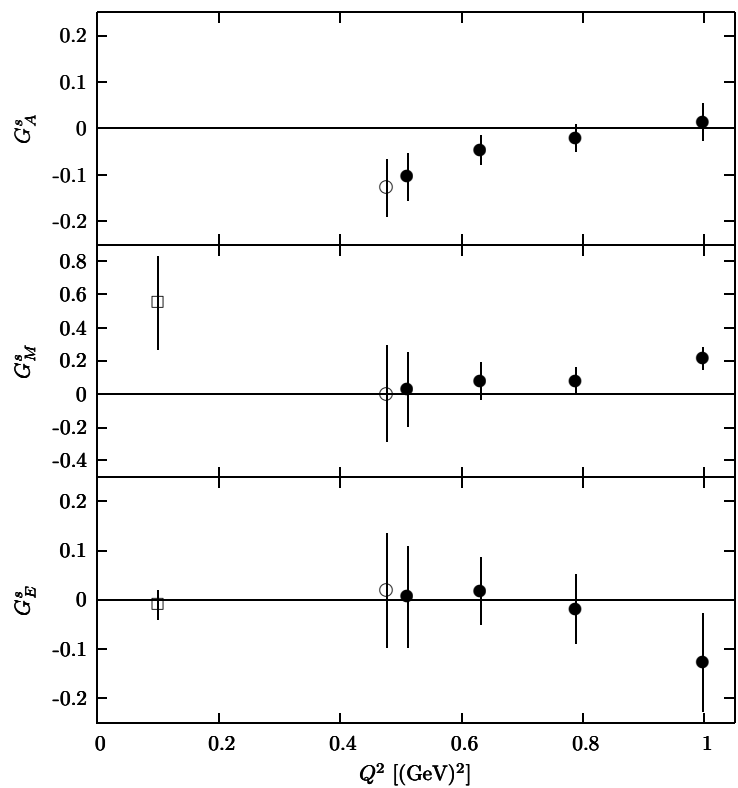}
  \caption{Results of this analysis:  Open circles are from a combination of 
HAPPEx and E734 data, while the
closed circles are from a combination of $G^0$ and E734 data.  [Open squares are
from Ref.~\cite{HAPPEx_1H_010} and involve PV $ep$ data only.]}
\label{sff_fig}
\end{figure}
{\bf Combining the two data sets:}
The basic technique for combining these two data sets has already been 
described~\cite{Pate:2003rk}
and the details of the present analysis will be published soon~\cite{PMMP}.
The results are displayed in Figure~\ref{sff_fig}.  The uncertainties
in all three form factors are dominated by the large uncertainties in the
neutrino cross section data.  Since those data are somewhat insensitive to
$G_E^s$ and $G_M^s$ then the uncertainties in those two form factors are
generally very large.  However the results for the strange axial factor 
are of sufficient precision to give a hint of the
$Q^2$-dependence of this important form factor for the very first time.
There is a strong indication from this $Q^2$-dependence that $\Delta s < 0$, {\em i.e.}
that the strange quark contribution to the proton spin is negative.  However
the data are not of sufficient quality to permit an extrapolation to
$Q^2=0$, so no quantitative evaluation of $\Delta s$ from these data can 
be made at this time.

A new experiment called FINeSSE~\cite{FINeSSE_FNAL_LOI} has been proposed
to measure the strange axial form factor to sufficient precision to
determine $\Delta s$, by measuring the ratio of the neutral-current
to the charged-current $\nu N$ and $\bar{\nu}N$ processes. A measurement of 
$R_{NC/CC}=\sigma(\nu p\rightarrow\nu p)/\sigma(\nu n\rightarrow\mu^- p)$
and
$\bar{R}_{NC/CC}=\sigma(\bar{\nu}p\rightarrow\bar{\nu}p)/\sigma(\bar{\nu}p\rightarrow\mu^+n)$
combined with the world's data on forward-scattering PV $ep$ data can produce
a dense set of data points for $G_A^s$ in the range $0.25<Q^2<0.75$ GeV$^2$ 
with an uncertainty at each point of about $\pm 0.02$, resulting in the first
form factor extraction of $\Delta s$.

\begin{theacknowledgments}
The authors are grateful to E. Beise and J. Arvieux for useful discussions.
This work was supported by the US Department of Energy.
\end{theacknowledgments}

\bibliographystyle{aipproc}   % if natbib is available

\bibliography{my_proc}

%%%%%%%%%%%%%%%%%%%%%%%%%%%%%%%%%%%%%%%%%%%
%% Just a reminder that you may have to run bibtex
%% All of it up to \end{document} can be removed
%% if you don't like the warning.
%%%%%%%%%%%%%%%%%%%%%%%%%%%%%%%%%%%%%%%%%%%
\IfFileExists{\jobname.bbl}{}
 {\typeout{}
  \typeout{******************************************}
  \typeout{** Please run "bibtex \jobname" to optain}
  \typeout{** the bibliography and then re-run LaTeX}
  \typeout{** twice to fix the references!}
  \typeout{******************************************}
  \typeout{}
 }

\end{document}